\documentclass[12pt,preprint]{aastex}

\def\dnu{$\Delta\nu$}
\def\numax{$\nu_{\rm max}$}
\def\teff{$T_{\rm eff}$}
\def\logg{$\log g$}
\def\feh{[Fe/H]}
\def\msun{M$_\odot$}

\shorttitle{Grid based asteroseismology}
\shortauthors{Basu et al.}

\begin{document}

\title{Effect of uncertainties in stellar model parameters on
estimated masses and radii of single stars}

\author{Sarbani Basu\altaffilmark{1}, Graham
A. Verner\altaffilmark{2,3}, William J. Chaplin\altaffilmark{2}
and Yvonne Elsworth\altaffilmark{2}}

\altaffiltext{2}{Department of Astronomy, Yale University, P.O. Box
208101, New Haven, CT 06520-8101, USA}
\email{sarbani.basu@yale.edu}

\altaffiltext{2}{School of Physics and Astronomy, University of
Birmingham, Edgbaston, Birmingham B15 2TT, UK; w.j.chaplin@bham.ac.uk,
y.p.elsworth@bham.ac.uk, miglioa@bison.ph.bham.ac.uk}

\altaffiltext{3}{Astronomy Unit, Queen Mary, University of London,
 Mile End Road, London E1 4NS, UK}

\begin{abstract}

Accurate and precise values of radii and masses of stars are needed to
correctly estimate properties of extrasolar planets. We examine the
effect of uncertainties in stellar model parameters on estimates of
the masses, radii and average densities of solar-type stars. We find
that in the absence of seismic data on solar-like oscillations,
stellar masses can be determined to a greater accuracy than either
stellar radii or densities; but to get reasonably accurate results the
effective temperature, \logg\ and metallicity must be measured to high
precision.  When seismic data are available, stellar density is the
most well determined property, followed by radius, with mass the least
well determined property.  Uncertainties in stellar convection,
quantified in terms of uncertainties in the value of the mixing length
parameter, cause the most significant errors in the estimates of
stellar properties.

\end{abstract}


\section{Introduction}
\label{sec:intro}

It is extremely difficult to estimate masses and radii of stars, in
particular masses of single stars. However, in this era of exoplanet
studies the masses and radii of exoplanet hosts play an important role
in establishing the properties of the planets.

The usual way to determine stellar properties is through spectroscopic
or photometric analyses of the star combined with fitting to grids of
stellar models (see e.g., Takeda et al. 2007). The recent, dramatic
advances in observational asteroseismology provide us with a
relatively new way to determine masses and radii of solar-type
stars. These advances have come in large part from new satellite
observations, for example from the French-led CoRoT satellite (e.g.,
Michel et al. 2008; Appourchaux et al. 2008), and in particular the
NASA \textit{Kepler} Mission (Chaplin et al. 2010; Gilliland et
al. 2010). \textit{Kepler} has provided detections of solar-like
oscillations for hundreds of solar-type stars (Chaplin et al. 2011)
when only around twenty such stars had detections prior to the
Mission.

The {\it Kepler} Mission (Borucki et al. 2010) is currently the
largest source of data on extra-solar planet candidates and their host
stars.  Properties of stars in the {\it Kepler} field of view are
available in the {\it Kepler} Input Catalog (KIC; Brown et al. 2011).
The catalog provides estimates of \teff, \logg, \feh, and $E(B-V)$
obtained from the analysis of multi-color photometry and also derived
properties such as radius, mass and luminosity.  Verner et al.~(2011)
compared the seismically derived radii of a sample of solar-type stars
with those in the KIC, and found that although there was general
agreement, there was an underestimation bias of up to 50\% for stars
with $R_{\rm KIC} < 2 R_\odot$, showing that estimates of stellar
properties are prone to uncertainties.

Stellar masses and radii obtained by modelling observed stellar
parameters such as the effective temperature \teff, metallicity \feh\
and surface gravity \logg\ are subject to systematic errors caused by
uncertainties in the inputs to stellar models, as well as errors that
arise because of uncertainties in the measured stellar parameters.  In
this paper we determine the effect of uncertainties in model
parameters on derived stellar masses, radii and average stellar
densities. The paper is organized as follows: in \S~\ref{sec:pot} we
discuss some possible sources of error; in \S~\ref{sec:prop} we
describe how we estimate stellar masses and radii; in
\S~\ref{sec:results} we describe the proxy stars that we use to
determine the effects of model uncertainties on estimated stellar
parameters. Our results are presented and discussed in
\S~\ref{sec:results} and we present the main conclusions in
\S~\ref{sec:conc}.

\section{Potential sources of error}
\label{sec:pot}

Stellar properties estimated from observations of temperature,
metallicity and gravity depend on stellar models. As a result, all
uncertainties in stellar models will affect the estimated mass and
radius of the star under consideration.  We consider three main
sources of uncertainty: (1) the mixing length parameter; (2) the
metallicity scale; and (3) the atmospheric model. Additionally, we
examine what happens if there are systematic errors in the temperature
scale.

One of the largest sources of uncertainty in stellar models is the
treatment of convection. Given that it is not possible to follow both
convective and evolutionary time-scales in a stellar model, convection
in stars is usually modeled using the so-called mixing length theory
(MLT) or one of its variants. Unfortunately, there is a free
parameter, the so-called mixing length parameter $\alpha$, that must
be specified before one can begin modeling.  The usual practice is to
assume the solar mixing length applies to all stars. The solar mixing
length (and the initial solar helium abundance) is determined by
forcing a 1\msun\ model to have the solar radius and solar luminosity
at the current solar age. While this is the most commonly used
approach, efforts to model stellar pulsation data shows that stars do
not necessarily have the solar value of $\alpha$ (e.g., see Metcalfe
et al. 2010; Benomar et al. 2010, Deheuvels \& Michel 2011). Thus,
assuming that the stars being analyzed have the solar value of
$\alpha$ will necessarily affect our results, particularly since
$\alpha$ controls the radii of stellar models.

Another potential source of error is the metallicity scale. Most
stellar metallicity estimates are given with respect to solar
metallicity. Converting to $Z$ hence requires a value of the solar
metallicity. Asplund, Grevesse \& Sauval (2005; henceforth AGS) had
suggested that $Z/X$ for the Sun is 0.0165, this value is much lower
than the usually accepted value of 0.023 (Grevesse \& Sauval 1998) or
0.0245 (Grevesse \& Noels 1993). Although Asplund et al.~(2009) and
Grevesse et al.~(2010) revised the AGS results upwards to
$Z/X=0.0181$, this is still lower than the Grevesse \& Sauval (1998)
value. This uncertainty in the solar metallicity scale adds
uncertainties in evolutionary tracks and hence needs examining.

The atmospheric $T-\tau$ relations used in stellar models also play a
role in defining the fundamental properties of models. It is common to
use the Eddington $T-\tau$ relation to define the outer boundary of
the star, but often other, semi-analytic $T-\tau$ relations, such as
those of Vernazza et al. (1981) or Krishna Swamy (1966), are
used. Since the $T-\tau$ relation effectively determines the layer at
which $T$=\teff, it plays a role in determining the radius of a model
of a given mass and temperature, and can add to uncertainties to the
estimated stellar parameters.

While the factors mentioned above increase uncertainties in stellar
models, one observational issue that can affect stellar parameter
determination is \teff. Determinations of \teff\ of the same star by
different groups often result in vastly different temperature
estimates (see e.g., Metcalfe et al.  2010 for the case of one of the
{\it Kepler} stars). There are often scale errors between one set of
photometric temperature differences and another with the differences
often being as large at 100K (e.g., see Casagrande et al. 2010;
Pinsonneault et al. 2011). This will affect estimates of stellar
parameters, and hence needs to be accounted for in the error-budget.

\section{Determining Stellar Properties}
\label{sec:prop}

Basu et al.~(2010) described a way (the Yale-Birmingham or YB code) in
which stellar radii can be determined using seismic parameters. The YB
pipeline was modified by Gai et al.~(2010) to obtain other stellar
properties, such as mass and \logg. Here, we use the same method for
determining masses and radii of stars even in the absence of seismic
data.

The YB method is based on finding the maximum likelihood of the set of
input parameter data calculated with respect to the grid models.  The
estimate of the stellar property is obtained by taking an average of
the properties on the grid that have the highest likelihood. We
average all values with likelihoods over 95\% of the maximum value of
the respective likelihood function.  For a given observational
(central) input parameter set, the first key step in the method is to
generate 10,000 input parameter sets by adding different random
realizations of Gaussian noise to the actual (central) observational
input parameter set. The distribution of any property, say radius, is
then obtained from the central parameter set and the 10,000 perturbed
parameter sets, which forms the distribution function. The final
estimate of the property is the median of this distribution.  We use
1$\sigma$ limits from the median as a measure of the uncertainties.

The likelihood function is formally defined as
 \begin{equation}
 \mathcal {L}=\prod^{n}_{i=1}\left(\frac{1}{\sqrt{2\pi}\sigma_{i}}\times
 \exp(-\chi^{2}/2)\right), \label{eq:likelihood}
 \end{equation}
where
 \begin{equation}
 \chi^{2}=\sum^{n}_{i=1}(\frac{q^{obs}_{i}-q^{model}_{i}}{\sigma^{i}})^{2},
 \label{eq:chi2}
 \end{equation}
with $q$ $\equiv$ \teff, \feh, and \logg\ in the non-seismic case, and
\teff, \feh, the average large frequency separation, \dnu, and the
frequency of maximum power, \numax, in the seismic case.

The two seismic observables, \dnu\ and \numax\, depend on global properties of a
star.  The large separation scales as the mean density of
a star (e.g., Ulrich 1986, Christensen-Dalsgaard 1993), so that
 \begin{equation}
 \frac{\Delta\nu}{\Delta\nu_{\odot}}\simeq\sqrt{\frac{M/M_{\odot}}{(R/R_{\odot})^{3}}}.
 \label{eq:delnu}
 \end{equation}
Thus, assuming we know the \dnu\ for the Sun, we can find the mean
density of other stars. Stello et al. (2009) have shown that this
scaling holds over most of the HR diagram and errors are probably
below 1\%. The frequency of maximum power in the oscillations power
spectrum, \numax, is related to the acoustic cut-off frequency of a
star (e.g., see Kjeldsen \& Bedding 1995; Bedding \& Kjeldsen 2003;
Chaplin et al. 2008), which in turn scales as $M\,R^{-2}\,T_{\rm
eff}^{-1/2}$. Thus, if we know the solar values of \numax, we can
calculate \numax\ for any star as:
 \begin{equation}
 {{\nu_{\rm max}}\over{\nu_{\rm max,\odot}}}\simeq{{M/M_{\odot}}\over
 {(R/R_{\odot})^2\sqrt{(T_{\rm eff}/T_{{\rm eff},\odot})}}}
 \label{eq:numax} 
 \end{equation}

If \dnu, \numax\ and \teff are known, Equations~(\ref{eq:delnu}) and
(\ref{eq:numax}) represent two equations in two unknowns, $M$ and $R$,
which hence can be solved to obtain the mass and radius. This also 
allows us to calculate \logg. 

Although Eqs.~\ref{eq:delnu} and \ref{eq:numax} can be used to
determine mass and radius directly, the errors in the derived
quantities can be unphysically large, in the sense that the equations
assume that all values of \teff\ are possible for a star of a given
mass and radius. However, the equations of stellar structure and
evolution tell us otherwise --- we know that for a given mass and
radius, only a narrow range of temperatures are allowed.  This is what
led to the development of so-called `grid-based' methods, such as the
YB pipeline, where the seismic observations are used in conjunction
with a grid of models. This does however come at a cost, since it
introduces some model dependence in the seismic estimates of the
stellar properties.

We use four grids of models to determine stellar parameters. These
``calibration'' grids are: (i) a grid of models that constitute the YY
isochrones (Demarque et al. 2004); (ii) a grid of models constructed
with the Dartmouth Stellar Evolution code (DSEP; Dotter et al. 2007),
as described by Dotter et al.~(2008) [these models can be downloaded
from the DSEP
webpage\footnote{http://stellar.dartmouth.edu/\~{}models/index.html}];
(iii) the model grid of Marigo et al.~(2008) constructed with the
Padova stellar evolution code (Marigo et al. 2008; Girardi et
al. 2000) [these models can be downloaded from the Padova CMD
webpage\footnote{http://stev.oapd.inaf.it/cgi-bin/cmd}]; and (iv) a
grid of models constructed with the Yale Rotation and Evolution Code
(YREC; Demarque et al.~2008) in its non-rotating configuration [these
models are described in Gai et al.~(2011)].  All  models have
been constructed with the solar-calibrated value of the mixing length
for the particular code and physics used. Although the input physics
in these grids is different, they are not sufficiently different from
each other for us to use models from these grids as test
models. Consequently we refer to models in these grids as ``normal
models''.

As can be seen from Eq.~\ref{eq:likelihood}, knowing the uncertainty
in the observations is key.  We use three sets of errors in this
work. Set 1 corresponds to the uncertainties believed to be present in
KIC for \teff, \feh\ and \logg.  Set 2 corresponds to typical errors
expected in (good) photometric estimates of \teff, \feh\ and
\logg. Set 3 is what is expected from spectroscopic
determinations. The uncertainties used are listed in
Table~\ref{tab:err}.  As can be seen in the table, for all three sets
we assume identical errors in the seismic properties.

Gai et al.~(2011) have already performed an in-depth analysis of
asteroseismic radius and mass determinations. However, they did not
consider all the sources of error that we shall consider here,
consequently we also examine how the above sources of error affect
seismic estimates of the stellar properties. For the seismic
determinations we will use the same inputs as in Gai et al.~(2011),
i.e., \teff, \feh, the large separation \dnu, and the frequency of
maximum power, \numax.

\begin{figure*}
\plottwo{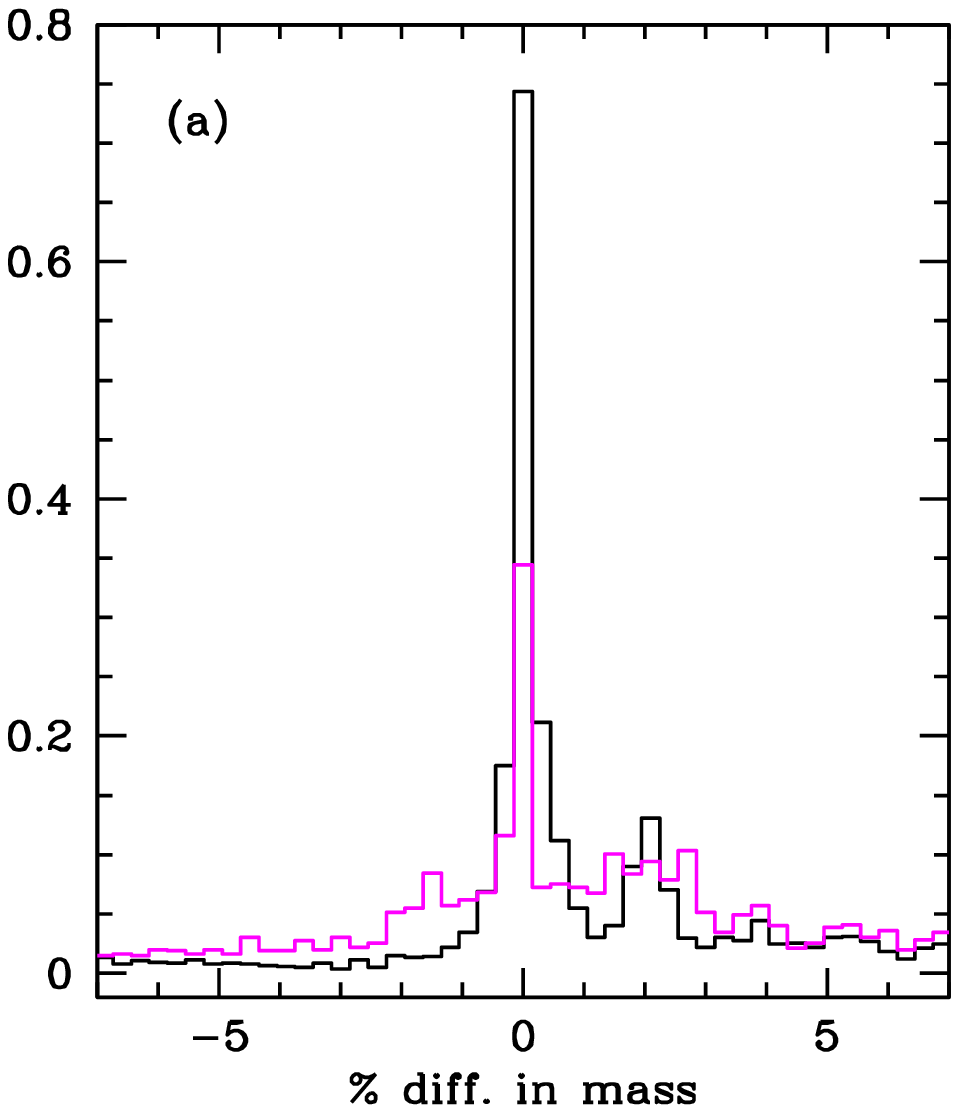}{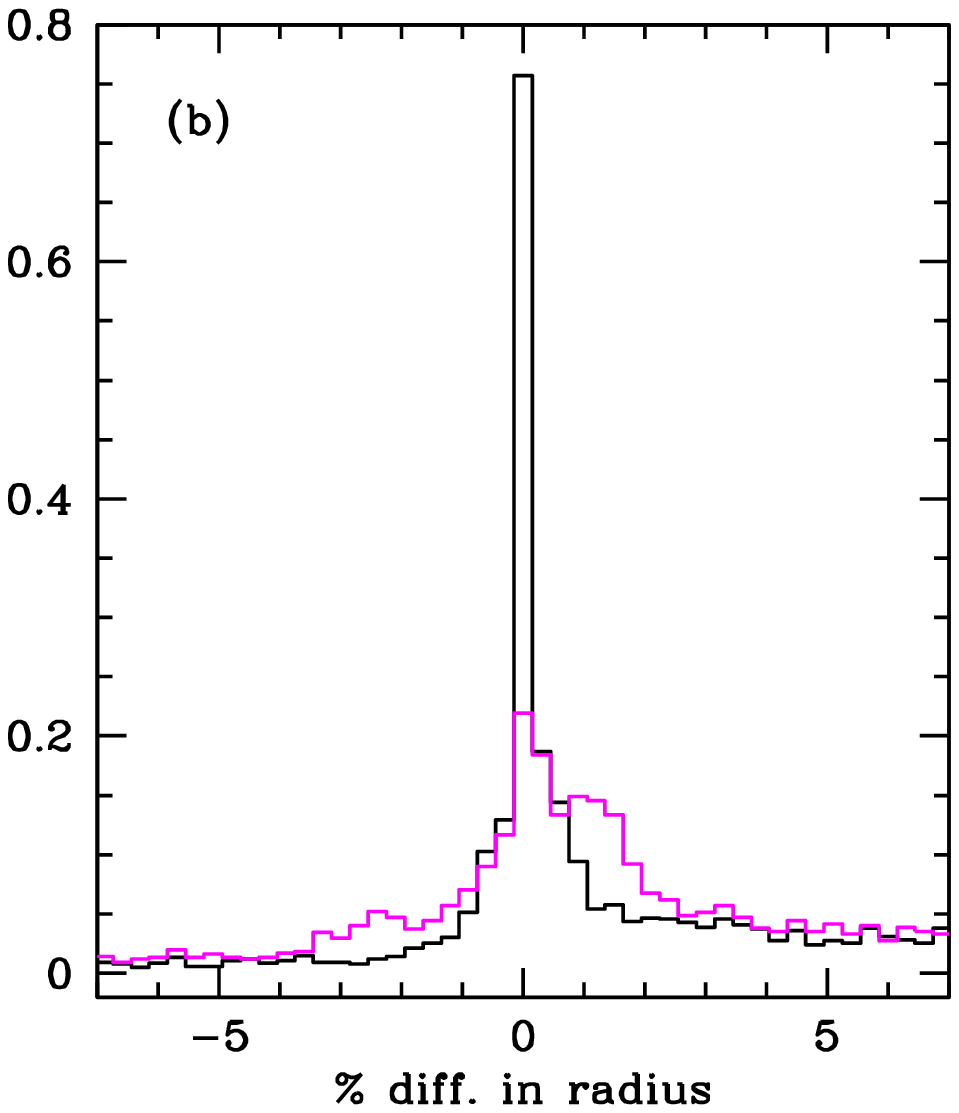}
\caption{The distributions, normalized to unit area, of the fractional
differences between the estimated and true masses (panel a) and radii
(panel b) using \teff, \feh, and \logg as inputs. The proxy stars were
derived from the same grid of stars used to estimate the stellar
properties. The black curve shows the results of using error-free
data, under the assumption that the input observations have no
errors. The magenta curve was obtained when we used error-free data
but assumed that the input parameters had errors as per the Error
case~3 (Table~\ref{tab:err}).}
\label{fig:nonseis}
\end{figure*}

Basu et al.~(2010) and Gai et al.~(2011) include extensive discussions
on how well the method works when seismic data are available. We find
that the method works even for the combination of non-seismic data
\{\teff, \feh, \logg\}, if we use synthetic stars drawn from the same
family of models as the grid. In the the absence of uncertainties in
any of the parameters, the half-width at half-maximum (HWHM) for the
distribution of differences between input and output masses and radii
is less than 0.25\%, while the semi-quartile distance (i.e., half the
distance between the 1st and 3rd quartile points) is about 2\%.  If we
use exact, error-free, data but assume that there could be errors as
in Error~Set~3, then in the case of mass the HWHM is still less than
0.25\%, but the semi-quartile distance is somewhat larger, at
2.8\%. For radius, the HWHM is 1\%, while the semi-quartile distance
is 7\%.  The probability distribution functions are shown in
Figure~\ref{fig:nonseis}.  The results show that we can use the same
technique for determining radii and masses of stars regardless of
whether or not we use seismic data.

\section{The Test Cases}
\label{sec:test}

In order to investigate the effects of the input parameters we have
used several sets of simulated stars drawn from different grids of
stellar models to determine the errors made in determining the masses,
radii and average densities of these stars.

In the first set of tests -- which we call the ``Identical'' set --
the proxy stars come from the same grid of models that is used to
estimate their properties, an important implication being that both
have the same physics. In the second set -- the so-called ``Normal''
set -- a mixture of proxy stars is drawn from all four grids, with the
name reflecting the fact that these grids were constructed with usual
stellar parameters. However, there are differences in detail between
the different grids (i.e., differences in the physics inputs used).

We also use four other sets of proxy stars derived from grids
constructed especially for this work. The first of these, ``AGS'',
assumes that the [Fe/H]=0 corresponds to $Z/X=0.0165$, the solar
metallicity claimed by Asplund et al.~(2005). This set should simulate
the effects of getting the metallicity scale wrong (since the grids
used to estimate the stellar properties assume the higher solar
$Z/X=0.023$, or similar). The AGS proxy stars have solar calibrated
values of the mixing length parameter.  The next set, ``MLT'',
comprises models constructed with non-solar values of the mixing
length, in particular we use models with $\alpha=1.5$, 1.6, 1.7, 1.9,
2.0. 2.1, 2.2, 2.3, and 2.4. The solar $\alpha$, for the same physics,
is $1.826$.  A third set of so-called ``KS'' models was constructed
with the Krishna Swamy (1966) $T-\tau$ relation in the atmosphere.  The
last set, which we call ``T-shift'', was derived from the four grids,
just like the ``Normal'' models; however, the effective temperatures
were increased by 100K to simulate the effect of getting the
temperature scale wrong. The different test cases are summarized in
Table~\ref{tab:test}.

In all cases, we added to the seismic and non-seismic input parameters of
the proxy stars random errors consistent with the uncertainties listed in
Table~\ref{tab:err}, and then used these perturbed data as the inputs
to our grid pipeline.

\section{Results and Discussion}
\label{sec:results}

Our aim is to see how well we recover the properties of the stars in
each of the test cases. To this end, we look at the fractional
deviations between the actual properties and the estimated properties
of the proxy stars. If there are no systematic effects, the
distribution of the fractional deviations will be symmetric and
centered at zero.  Our results for all test cases analyzed, and for
all three sets of errors, are presented in Table~\ref{tab:res}. In
each case we list the median of the distribution as well as the spread
(standard deviation of the distribution). 

These two values do not completely characterize the distributions and
many of them are highly skewed, as we shall see below, and hence, we
also list the HWHM in both the negative error and positive error
directions. The spread as well as the two HWHMs give information on
the non-Gaussianity as well as the asymmetry of the distributions.
The data in the table allow us to draw some important conclusions: (1)
the errors in the results, as expected, decrease when the
uncertainties in the simulated input data are decreased; (2) for the
same set of errors, adding seismic inputs improves the results
considerably; and (3) differences in physics (i.e., between the proxy
stars and the model grids used to estimate the properties) lead to
noticeable biases in the stellar properties estimated from non-seismic
inputs alone, while the effects are much reduced when seismic input
data are available.

\begin{figure*}
\plotone{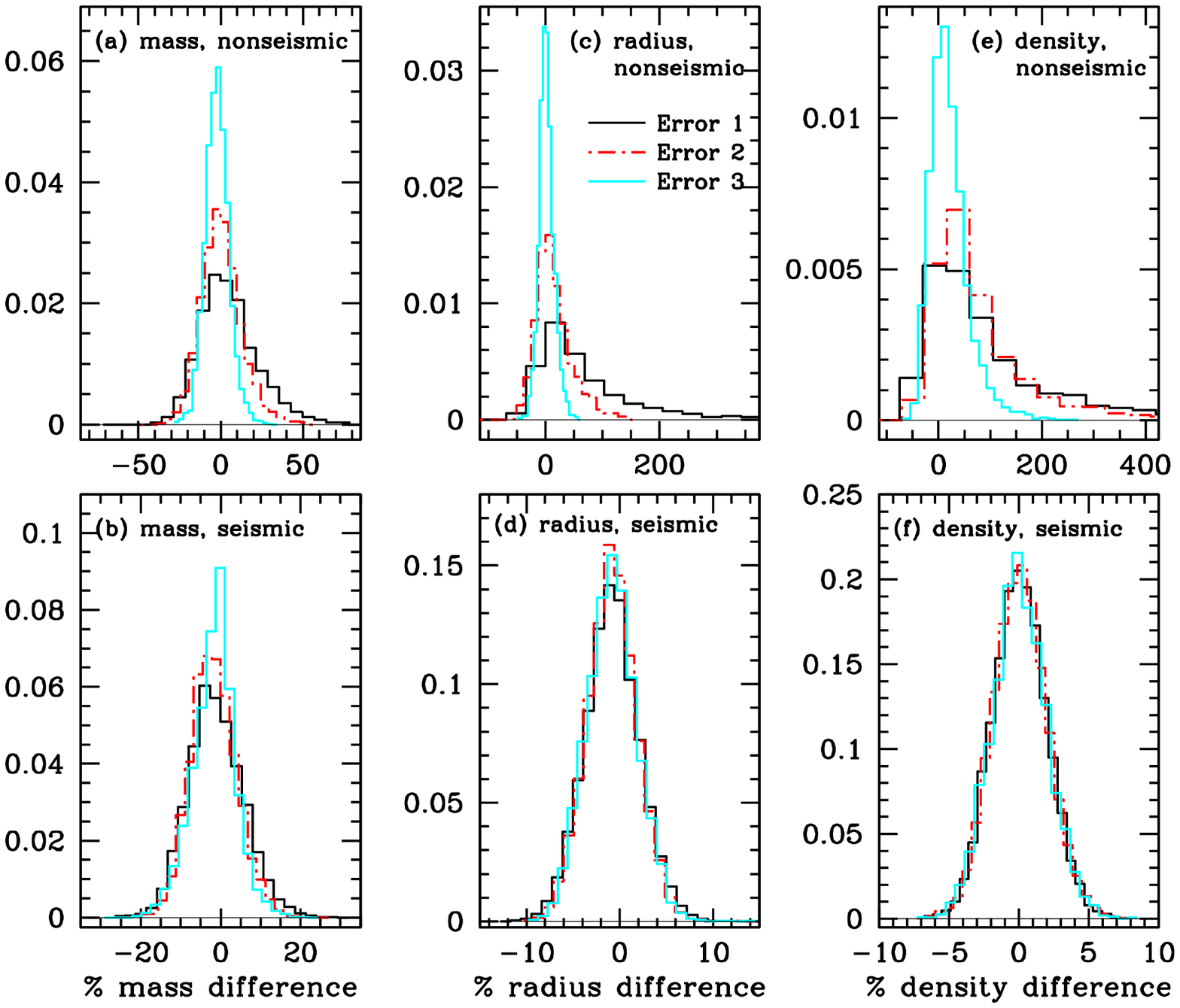}
\caption{The distributions, normalized to unit area, of the fractional
differences between the estimated and true masses, radii and densities
of the ``Normal'' proxy stars, for all three error cases. The upper
panels show the distributions obtained without the use of seismic
data, while the lower panels show those obtained when seismic data are
used.}
\label{fig:normal}
\end{figure*}

The effect of errors is shown in Fig.~\ref{fig:normal}, where we plot
the distribution of the errors for the Normal set results (all three
error cases). In the non-seismic case, the best-determined property is
the mass. This is not surprising since stellar model properties depend
predominantly on (or are fixed predominantly by) mass.  As can be seen
from the figure, decreasing the input parameter uncertainties makes a
dramatic difference in the non-seismic results: unless we have very
precise estimates of \teff, \feh\ and \logg, mass errors can be as
much as 50\% in some cases. Radius determinations are much more
uncertain, with density faring the worst (since the mass and radius
errors propagate through). In our best-case scenario, i.e., Error case
3, the half-width at half maxima for mass, radius and density are 8\%,
12\% and 35\%, respectively.

In contrast, the best-determined quantity in the seismic case is
density, which is not surprising since the input \dnu\ scales as
$\sqrt{\rho}$. The errors in the estimated densities are almost
completely independent of the errors in the other input parameters,
the error-distributions being consistent with a Gaussian with
$\sigma=2$\% (i.e., entirely consistent with the 1\% error adopted for
\dnu).  The next-best determined property is radius. Here, errors in
\teff\ and metallicity play only a minor role, as reported by Gai et
al.~(2011).  In the seismic case, mass determinations fare the worst
and reducing uncertainties in \teff\ and \feh\ is essential. However,
the mass results are still better than the non-seismic case with HWHM
of 8\%, 7\%, 5\% for Error cases 1, 2 and 3, respectively.

\begin{figure*}
\plotone{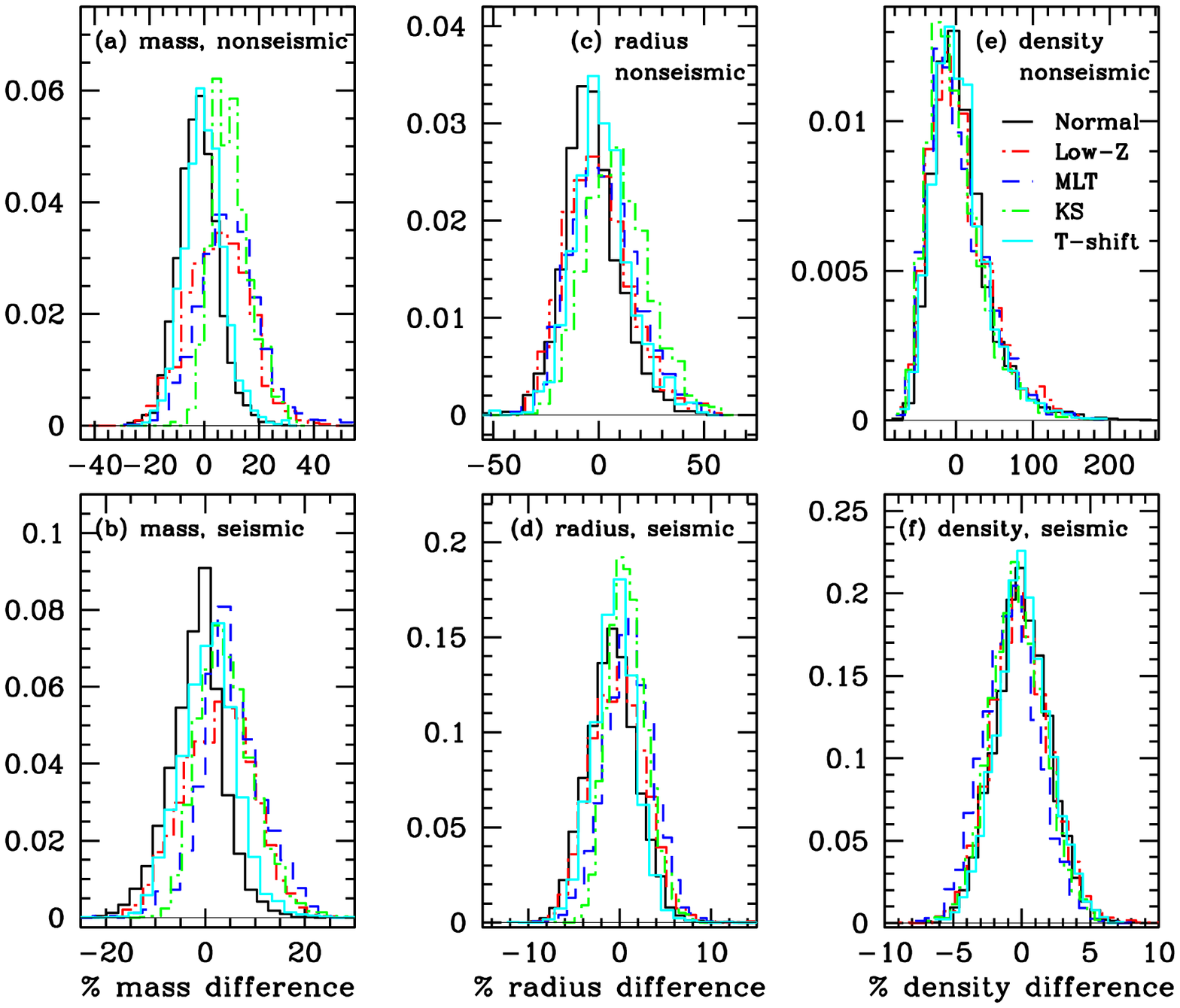}
\caption{The distributions, normalized to unit area, of the fractional
differences between the estimated and true masses, radii and densities
of the different sets of proxy stars, for Error case~3.  The upper
panels show the distributions obtained without the use of seismic
data, while the lower panels show those obtained when seismic data are
used.}
\label{fig:all}
\end{figure*}

Figure~\ref{fig:all} shows the error-distribution when the other sets
of proxy stars are used.  We only show the distributions for Error
case~3 and have also plotted the ``Normal'' set results as a
reference. What is clear from the figure is that there is a systematic
shift in the results when the properties of the other proxy stars are
deduced using normal stellar grids.  The median value of the
distributions listed in Table~\ref{tab:res} attest to the shifts.  We
have not shown the results when the proxy stars are identical to the
grid models; the results for identical stars are however listed in
Table~\ref{tab:res}, and the spread of the results is actually a
measure of the precision of the results. The spread of the other cases
is, in some sense, a convolution of the precision and the accuracy of
the method. While the results are equally precise in all cases, they
are not as accurate, which gives rise to a shift of the peak of the
distributions.

The effect of uncertainties in the physics is most noticeable in the
mass estimates, for both non-seismic and seismic cases.  The peak of
the distribution shifts away from zero and in many cases the
distribution becomes wider.  The effects on radius are smaller in a
relative sense, and the effects on density are smaller again, relative
to what we had seen in the Normal case. Given that mass is believed to
be the most fundamental property of a star, and one that controls the
evolution, this is somewhat disappointing although not very difficult
to explain. In the absence of seismic data, the only observable that
depends explicitly on mass is \logg. Although the observed \teff\ is
determined by mass and metallicity, its explicit dependence is on
radius.  In the seismic case too, the \dnu\ and \numax\ parameters
depend more on radius than on mass. To determine masses of single
stars better, we need to model the full oscillating spectrum of a
star, not merely \dnu\ and \numax\ (see e.g., Metcalfe et al. 2010,
Deheuvels \& Michel 2011).

Of all cases, an artificial shift in the \teff\ scale produces the
smallest effect. While the peak of the distribution shifts slightly,
the width does not change for the non-seismic estimates, while the
HWHM increases modestly, to only 6\% (from 5\%), for the seismic
estimates. The relative insensitivity to the temperature scale is
reassuring since it is not completely clear how the temperature scale
can be improved in the short term. Of course, the result would be
different if we shifted the temperatures by a much larger amount.

Differences in other physics inputs cause large changes, especially in
the absence of seismic data. We find that masses of the Low-Z proxy
stars are generally overestimated by our calibration grids.  One
simple way to understand this is the following: Masses of stars for
which [Fe/H]$=0$ is pegged to the low value of solar $Z/X$ are
overestimated with grids of normal models. A low $(Z/X)$ for a given
helium abundance results in a star that is hotter than a higher $Z/X$
star. For a given \logg, these models have a higher \teff, mimicking a
higher-mass star in the grid of models that has an [Fe/H] scale pegged
to the higher $Z/X$. Since the latest set of abundances published by
Asplund et al.~(2009) and Grevesse et al.~(2010) are higher than those
of AGS (although still lower than the GS98 value), the errors we get
for the ``AGS'' set of proxy stars could be considered to be an upper
limit to the errors caused by the metallicity scale.

Unsurprisingly, proxy stars with non-solar mixing length also show
systematic errors.  Basu et al.~(2010) had shown that even when
seismic data are available, uncertainties in the mixing length can
result in systematic errors in radius measurements. We find that in
the absence of seismic data, the errors are much larger, particularly
in the case of mass estimates.  For a given mass, a change in $\alpha$
leads to a change in radius, with sub-solar $\alpha$ models showing a
larger radius (and hence lower \teff\ and \logg) at a given luminosity
than solar $\alpha$ models with the same metallicity. Super-solar
$\alpha$ models show a smaller radius (and hence larger \teff\ and
\logg). In most parts of the HR diagram this change in \logg\ (at a
given \teff) seems to be interpreted more like a change in mass than
the change in radius that it actually is, giving rise to errors in
mass estimates that can be quite large. There are errors in the radius
estimation too, but the resulting errors, relative to the errors in
the case of normal models, are not large.

While estimates of mass and radius that are made when seismic data are
available also suffer from errors due to uncertainties in inputs to
the stellar models, the effects are much smaller. The main reason for
this is that the seismic input parameters depend directly on both
radius and (albeit to a lesser extent) on mass.

\begin{figure*}
\plotone{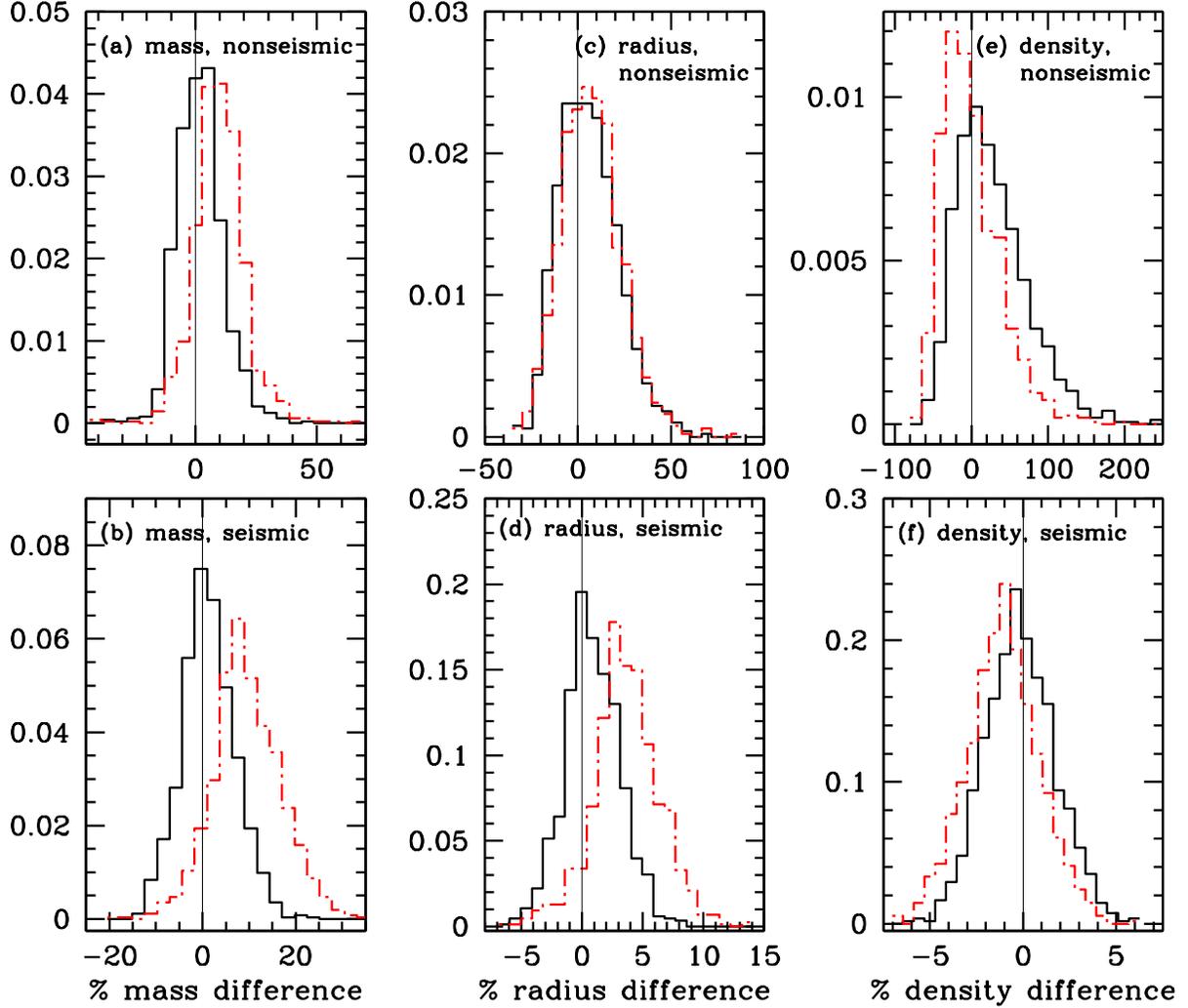}
\caption{The distributions, normalized to unit area, of the fractional
differences between the estimated and true masses, radii and densities
of the `MLT' set of proxy stars, for Error case~3. The red dot-dashed
histogram is the result of using the YREC grid, the black histogram is
the result of using a multi-$\alpha$ grid.
}
\label{fig:mlt}
\end{figure*}

The results for non-solar mixing length stars can be improved by using
a grid that consists of several complete grids, each with a different
mixing length. To test this we constructed YREC grids with five values
of $\alpha$, $\alpha=$ 1.4, 1.6, 1.826 (solar), 2.0 and 2.2 and
re-derived the properties of the MLT stars using this grid.  The
results are indeed improved compared with the single $\alpha$ YREC
grid, as can be seen from Figure~\ref{fig:mlt}, where we show results
for the Error~3 case. As can be seen from the figure, all
distributions are peaks at zero difference. Note that the
single-$\alpha$ results seem worse that those in Figure~\ref{fig:all};
this is because the results in Figure~\ref{fig:all} are a result of
using four different grids, each with a slightly different value of
$\alpha$ ($\alpha=1.938$ for Dotter et al., $\alpha=1.68$ for Marigo
et al., $\alpha=1.74$ for YY, and $\alpha=1.826$ for
YREC). Figure~\ref{fig:mlt} demonstrates the need for making grids
with a large range of values of the mixing length parameter, $\alpha$,
to obtain accurate properties of stars.

\begin{figure}
\plotone{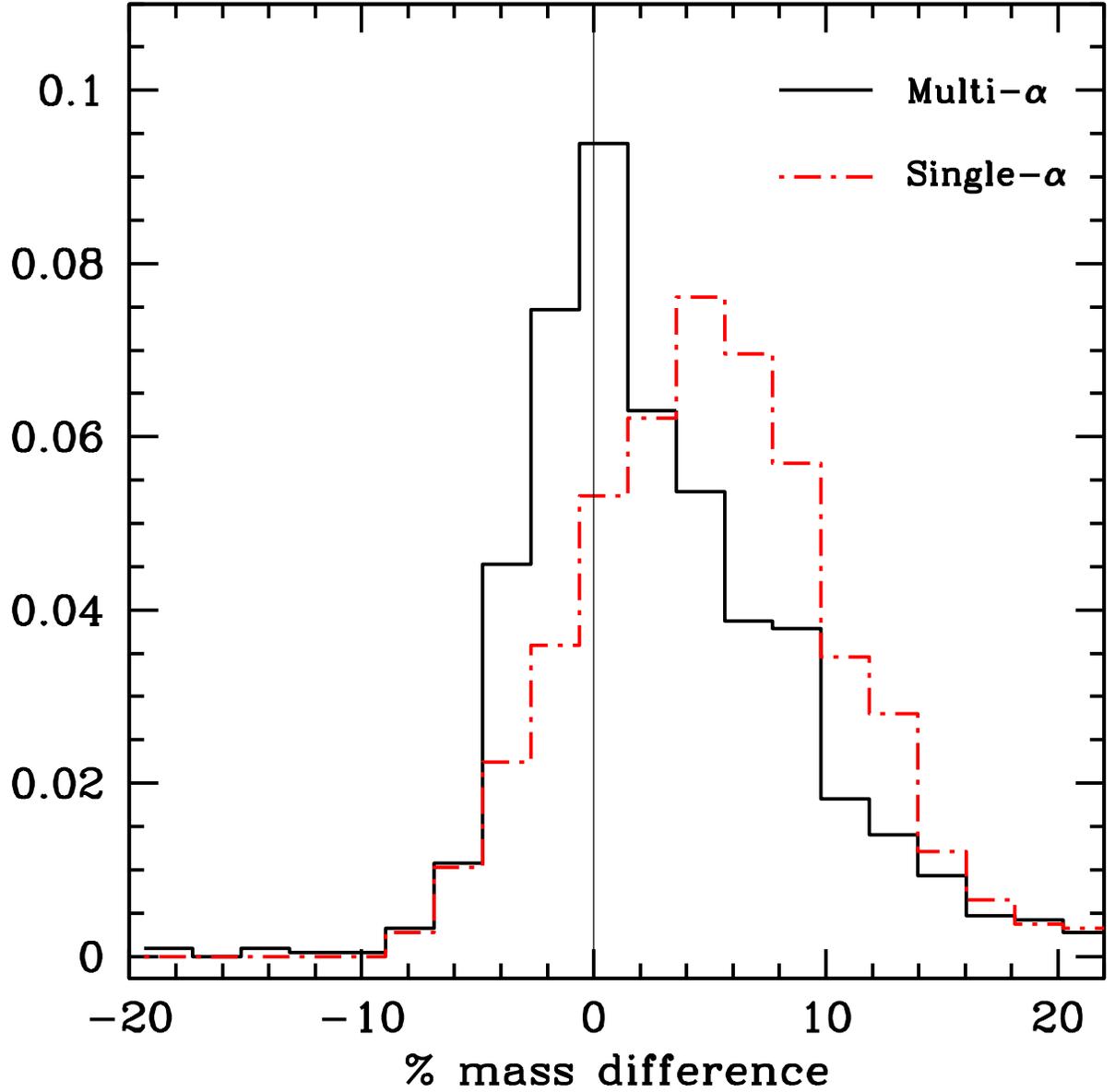}
\caption{The distribution, normalized to unit area, of the fractional
differences between the estimated and true masses
of the `KS' set of proxy stars (for Error case~3). The red dot-dashed
histogram is the result of using the YREC grid, the black histogram is
the result of using a multi-$\alpha$ grid. The results for the other quantities
are similar to those of the `MLT' set shown in Fig.~\ref{fig:mlt}.
}
\label{fig:ks}
\end{figure}

The results for ``KS'' proxy stars, made with the Krishna Swamy
atmospheres, can be understood to some extent in terms of the mixing
length. A solar calibrated model with a Krishna Swamy atmosphere needs
a mixing length parameter of $\alpha=2.1$. The proxy stars we used
were constructed with $\alpha=1.826$, which is the value of $\alpha$
we need to construct a solar model when the Eddington $T$-$\tau$
relation is used. Thus, the ``KS'' models in effect have sub-solar
mixing length. This means that for the same mass and metallicity, the
``KS'' proxy stars have a higher \logg\ at a given temperature. As in
the case of the ``MLT'' proxy stars, this appears to be interpreted as
a change in mass. The ``KS'' results also improve if we use a
multi-$\alpha$ grid, as demonstrated in Figure~\ref{fig:ks},
confirming that the effect of having a different model of stellar
atmospheres is essentially the same as changing $\alpha$.

One might question whether it would be better in the seismic case to
use Eqs.~\ref{eq:delnu} and \ref{eq:numax} directly to estimate the
stellar properties, thereby removing the sensitivity of the results to
the stellar models (and the uncertainties which arise from choices
made concerning the input physics). While this will lead to more
accurate results, such a strategy comes at the cost of giving much
larger uncertainties (i.e., it reduces the precision in the estimated
properties), for the reasons mentioned in \S~\ref{sec:prop}, and shown
in Gai et al.~(2011). Our exercise does however indicate that unless
we use a grid constructed with a wide range of values of the mixing
length parameter, the formal uncertainties in the results only give a
lower bound to the error in the result. Even with a multi-$\alpha$
grid, we will need to make an allowance because of uncertainty in the
metallicity scale.

If we consider the complete ensemble of all proxy stars (but do not
include the ``T-shift'' stars), the HWHM for mass, radius and density
for the non-seismic estimates (Error case~3) are 12.5\%, 17.8\% and
36\%, respectively. For the analagous seismic cases, they are 7.6\%,
3.2\% and 2.3\%, respectively. Again, these errors can be decreased if
we use grids made to cover a large range in $\alpha$.

Estimates of density are robust in all cases (in the sense that the
deviation from the true density remains unchanged). Radius estimates
change only slightly in the seismic case (HWHM rising to 3.2\% from
2.5\%), but rise somewhat more in the non-seismic case. It should be
noted that the errors we find are in some sense a lower limit to the
possible errors, particularly since modelling convection in stars
remains uncertain. Additionally, we would expect all the sources of
error to be present together, making the situation more complicated.

\section{Conclusions}
\label{sec:conc}

Uncertainties in inputs to stellar models add uncertainties in
properties of single field stars estimated from observational input
parameters \teff, \logg, and \feh.  In the absence of seismic data,
extremely precise estimates of these parameters are needed to make
reasonable estimates of the masses and radii of the stars. With
\teff-errors of only 50\,K and errors of 0.1\,dex in [Fe/H] and \logg,
we cannot determine masses to better than 8\% and radii to better than
than 14\%, even if we assume that we know the physics of stars
perfectly. Note that we are talking about the accuracy of the results;
the precision is often better. Uncertainties in stellar model
parameters -- in particular uncertainties in the mixing length
parameter -- can increase those errors by more than a factor of
two. Uncertainty surrounding the solar metallicity (which establishes
the [Fe/H] scale) also increases the errors.  Small changes in the
effective-temperature scale do not affect the results substantially.
Although results obtained by having seismic input parameters data
available are also affected by uncertainties in stellar physics
inputs, the effects are much smaller. Mass estimates are affected the
most in the seismic case, with errors increasing to about $\pm7$\% for
the cases considered.  In the absence of seismic input data, the
average stellar density cannot be determined very well; however with
seismic data the average stellar density becomes the most well
determined property, with errors almost completely determined by
errors in the seismic parameter \dnu. We find that errors in mass and
radius estimates can be decreased if we use underlying grids
constructed with multiple values of the mixing length parameter.

\acknowledgements

This work is partially supported by NSF grant AST-1105930 to SB. WJC,
YE and GAV acknowledge financial support from the UK Science and
Technology Facilities Council (STFC).

\newpage

\clearpage

\begin{table}
\caption{Assumed errors}
\begin{tabular}{lccc}
\hline
Parameter &\multicolumn{3}{c}{1$\sigma$ error}\\
\hline
& Error 1 & Error 2 & Error 3\\
\hline
\teff\ (K)          & 200     & 100    & 50   \\
{[Fe/H]} (dex)        & 0.5     & 0.25   & 0.1   \\
\logg  (dex)        & 0.5     & 0.25   & 0.1   \\
\dnu                & 1\%     & 1\%    & 1\%   \\
\numax              & 2.5\%   & 2.5\%  & 2.5\%   \\
\hline
\end{tabular}
\label{tab:err}
\end{table}

\clearpage

\begin{table}
\caption{Description of the sets of proxy stars}
\begin{tabular}{lcl}
\hline
Name      &&      Description\\
\hline
Identical &&     Subset of YREC models analyzed only with YREC grid.\\
Normal    &&     Subset of models taken from the YY, YREC, Dotter et al. and Marigo et al. grids. \\
          &&     [Fe/H] for the models is calculated assuming [Fe/H]$ =0 \equiv Z/X=0.023$.\\
Low-Z     &&     A subset of the normal models, but with [Fe/H] calculated assuming\\
          &&     [Fe/H] $= 0 \equiv Z/X=0.0165$.\\
MLT       &&     A set of models constructed with non-solar values of the\\
          &&     mixing length parameter $\alpha$. \\
          &&     In these models [Fe/H]$ =0 \equiv Z/X=0.023$.\\
KS        &&     A set of models constructed with Krishna Swamy (1966) model atmospheres.\\
          &&     In these models [Fe/H]$ =0 \equiv Z/X=0.023$.\\
T-Shift   &&     A subset of the `Normal' models, but with their \teff\ values assumed to\\
          &&     be 100K higher than the true \teff\ value.\\
\hline
\end{tabular}
\label{tab:test}
\end{table}

\clearpage
\begin{table*}
\caption{Errors in estimated radius, mass and density of synthetic stars. Results
are expressed as percentages.}
%{\scriptsize
{\tiny
\begin{tabular}{lrrrccrrrccrrrc}
\hline
 &\multicolumn{4}{c}{Error 1}& &\multicolumn{4}{c}{Error 2} & &\multicolumn{4}{c}{Error 3}\\
\hline
 & Median & Spread & $-$HWHM & $+$HWHM & & Median & Spread & $-$HWHM & $+$HWHM & & Median & Spread & $-$HWHM & $+$HWHM \\
\hline
\multicolumn{15}{c}{Mass (Nonseismic)}\\
\hline
Identical & $ 6.52$ & $11.57$  & $18.07$ & $14.23$   && $ 5.67$ &  $ 8.37$ & $11.12$  & $13.36$ &&  $ 3.27$ &  $ 5.42$ & $ 8.80$ & $ 6.78$\\
Normal    & $-0.46$ & $18.70$  & $12.66$ & $21.72$   && $-2.85$ &  $12.73$ & $11.37$  & $12.93$ &&  $-3.62$ &  $ 7.45$ & $ 8.61$ & $ 6.85$ \\
Low-Z     & $ 6.77$ & $23.78$  & $17.21$ & $24.05$   && $ 3.45$ &  $16.96$ & $ 8.13$  & $28.01$ &&  $ 3.40$ &  $11.04$ & $14.58$ & $13.56$ \\
MLT       & $ 9.74$ & $21.60$  & $16.23$ & $34.41$   && $ 8.53$ &  $16.21$ & $10.30$  & $25.34$ &&  $ 8.87$ &  $11.61$ & $ 9.41$ & $14.18$ \\
KS        & $32.24$ & $27.65$  & $ 9.97$ & $71.06$   && $20.09$ &  $18.76$ & $25.97$  & $18.52$ &&  $ 7.88$ &  $ 8.05$ & $12.40$ & $ 5.24$ \\
T-shift   & $ 1.55$ & $18.50$  & $10.53$ & $24.52$   && $-1.06$ &  $12.74$ & $ 9.88$  & $17.08$ &&  $-1.62$ &  $ 7.41$ & $ 6.17$ & $ 8.76$ \\
\hline
\multicolumn{15}{c}{Mass (Seismic)}\\
\hline
Identical &  $ 0.18$ & $4.14$ & $ 4.26$ & $ 9.51$  &&  $ 0.22$ & $5.66$  & $5.26$ & $ 4.74$ &&    $ 0.11$& $1.95$ & $2.55$ & $4.63$ \\
Normal    &  $-2.91$ & $6.98$ & $ 5.04$ & $10.44$  &&  $-3.12$ & $5.89$  & $5.41$ & $ 8.14$ &&    $ 1.31$& $4.66$ & $5.98$ & $3.32$ \\
Low-Z     &  $ 0.62$ & $6.62$ & $ 8.61$ & $10.57$  &&  $-0.04$ & $6.79$  & $6.71$ & $10.78$ &&    $-0.83$& $6.28$ & $8.55$ & $8.53$ \\
MLT       &  $ 2.06$ & $7.81$ & $10.09$ & $ 7.42$  &&  $ 3.50$ & $6.86$  & $6.41$ & $ 7.76$ &&    $-5.55$& $6.61$ & $4.41$ & $6.07$ \\
KS        &  $ 5.06$ & $7.05$ & $ 7.93$ & $ 4.80$  &&  $ 3.42$ & $6.07$  & $5.08$ & $ 6.09$ &&    $-1.58$& $4.96$ & $5.49$ & $6.84$ \\
T-shift   &  $-1.91$ & $6.86$ & $ 5.23$ & $10.33$  &&  $-1.90$ & $5.75$  & $7.00$ & $ 6.93$ &&    $ 1.49$& $4.77$ & $7.75$ & $4.00$ \\
\hline
\multicolumn{15}{c}{Radius (Nonseismic)}\\
\hline
Identical & $-11.89$& $ 31.89$ & $23.23$ & $ 45.18$ &&   $-2.65$ &  $17.83$ & $17.52$ & $30.08$  &&  $ 1.11$&  $ 9.58$ & $ 9.22$ & $19.06$ \\
Normal    & $ 23.02$& $ 89.93$ & $37.30$ & $ 53.34$ &&   $ 2.75$ &  $33.44$ & $26.71$ & $26.59$  &&  $-1.64$&  $13.47$ & $ 9.42$ & $14.65$ \\
Low-Z     & $ 22.48$& $ 96.63$ & $21.02$ & $113.47$ &&   $ 1.75$ &  $38.82$ & $23.72$ & $50.82$  &&  $ 1.32$&  $15.30$ & $16.52$ & $17.41$ \\
MLT       & $ 30.83$& $ 99.68$ & $57.19$ & $ 86.97$ &&   $ 5.90$ &  $39.33$ & $36.56$ & $35.49$  &&  $ 4.61$&  $16.01$ & $15.56$ & $22.29$ \\
KS        & $ 74.43$& $ 96.35$ & $55.75$ & $102.51$ &&   $38.35$ &  $49.73$ & $58.69$ & $48.91$  &&  $10.62$&  $15.27$ & $18.25$ & $15.95$ \\
T-shift   & $ 69.65$& $102.17$ & $58.57$ & $ 94.32$ &&   $21.06$ &  $40.07$ & $31.98$ & $35.10$  &&  $ 3.57$&  $13.63$ & $ 9.95$ & $14.80$\\
\hline
\multicolumn{15}{c}{Radius (Seismic)}\\
\hline
Identical &  $ 0.21$ & $1.85$ & $2.06$ & $3.81$ &&  $ 0.18$ & $1.64$ & $1.46$ & $2.84$ &&    $ 0.18$ & $1.66$ & $2.12$ & $2.88$ \\
Normal    &  $-0.96$ & $2.92$ & $2.85$ & $3.57$ &&  $-0.92$ & $2.48$ & $2.65$ & $3.14$ &&    $-0.91$ & $2.48$ & $3.14$ & $2.86$ \\
Low-Z     &  $ 0.11$ & $2.85$ & $3.58$ & $3.60$ &&  $-0.16$ & $2.58$ & $2.53$ & $4.04$ &&    $ 0.23$ & $2.45$ & $4.16$ & $3.19$ \\
MLT       &  $ 0.45$ & $2.99$ & $3.12$ & $4.45$ &&  $ 1.32$ & $2.64$ & $3.62$ & $1.81$ &&    $ 2.24$ & $2.44$ & $3.39$ & $2.48$ \\
KS        &  $ 1.53$ & $2.60$ & $2.24$ & $3.04$ &&  $ 1.16$ & $2.26$ & $1.98$ & $2.49$ &&    $ 0.76$ & $1.82$ & $1.87$ & $3.19$ \\
T-shift   &  $-0.88$ & $3.73$ & $2.53$ & $3.64$ &&  $-0.82$ & $3.48$ & $1.98$ & $3.21$ &&    $-0.68$ & $3.50$ & $3.12$ & $2.02$ \\
\hline
\multicolumn{15}{c}{Density (Nonseismic)}\\
\hline
Identical &$  3.94$ & $ 65.23$ & $40.24$ & $ 49.77$  &&  $ 5.18$&  $ 40.25$& $63.52$  & $58.79$ &&  $ 6.61$ &  $22.04$ & $34.36$ & $31.27$\\
Normal    &$ 10.76$ & $179.24$ & $31.24$ & $115.00$  &&  $ 6.53$&  $109.00$& $59.31$  & $53.77$ &&  $ -0.10$&  $41.37$ & $35.23$ & $34.45$\\
Low-Z     &$ 22.67$ & $224.86$ & $61.92$ & $ 70.08$  &&  $21.74$&  $130.50$& $57.11$  & $72.83$ &&  $ -1.70$&  $46.37$ & $30.44$ & $35.04$\\
MLT       &$ 18.29$ & $222.91$ & $48.78$ & $ 86.89$  &&  $-8.73$&  $ 50.06$& $23.95$  & $55.51$ &&  $ -4.85$&  $42.99$ & $22.63$ & $46.01$\\
KS        &$ 41.13$ & $317.34$ & $33.46$ & $ 93.54$  &&  $19.11$&  $152.12$& $45.95$  & $58.62$ &&  $ -7.80$&  $46.52$ & $19.29$ & $47.48$\\
T-shift   &$ 12.96$ & $186.18$ & $34.13$ & $103.26$  &&  $ 7.07$&  $108.45$& $41.35$  & $67.58$ &&  $ -0.10$&  $37.45$ & $28.27$ & $35.00$\\
\hline
\multicolumn{15}{c}{Density (Seismic)}\\
\hline
Identical & $-0.04$ & $1.93$ & $2.29$ & $2.41$  &&    $-0.03$ & $1.91$ & $2.01$ & $2.02$ &&    $-0.01$ & $1.89$ & $1.98$ & $2.27$ \\
Normal    & $ 0.06$ & $2.08$ & $2.22$ & $2.39$  &&    $ 0.05$ & $1.93$ & $2.46$ & $2.05$ &&    $ 0.00$ & $1.95$ & $1.93$ & $2.28$ \\
Low-Z     & $ 0.04$ & $1.95$ & $2.28$ & $2.24$  &&    $ 0.02$ & $2.06$ & $2.55$ & $2.01$ &&    $ 0.16$ & $2.21$ & $2.18$ & $2.53$ \\
MLT       & $-0.18$ & $1.93$ & $2.27$ & $2.33$  &&    $-0.37$ & $1.93$ & $1.95$ & $2.26$ &&    $-0.73$ & $2.00$ & $2.74$ & $1.92$ \\
KS        & $-0.19$ & $1.94$ & $2.86$ & $1.57$  &&    $-0.25$ & $1.92$ & $2.15$ & $2.36$ &&    $-0.35$ & $1.89$ & $2.04$ & $2.57$ \\
T-shift   & $ 0.18$ & $1.91$ & $2.12$ & $2.29$  &&    $ 0.16$ & $1.93$ & $2.43$ & $1.98$ &&    $ 0.07$ & $1.87$ & $1.83$ & $2.16$ \\
\hline

\end{tabular}
}
\label{tab:res}
\end{table*}


\begin{thebibliography}{}

\bibitem[Appourchaux et al.(2008)]{2008A&A...488..705A} Appourchaux,
T., Michel, E., Auvergne, M., et al.\ 2008, A\&A, 488, 705

\bibitem[Asplund et al.(2005)]{ags05}
Asplund, M.,  Grevese, N.,  Sauval, A.J., 2005, ASPCS, 336, 25

\bibitem[Asplund et al.(2009)]{ags09}
Asplund, M.,  Grevese, N.,  Sauval, A.J., Scott. P., 2009, ARA\&A, 47, 481

\bibitem[Baglin et al.(2002)]{bag02}
Baglin, A., Auvergne, M., Catala, C., Michel, E., Goupil, M. J., Samadi, R., Popielsky, B.,
2002, in proc. IAU Colloq. 185, ASPCS, 259, 625

\bibitem[Basu et al.(2010)]{bas10}
Basu, S. Chaplin, W. J., Elsworth, Y. 2010, ApJ, 710, 1596

\bibitem[Bedding \& Kjeldsen(2003)]{bedding03}
Bedding, T. R. \& Kjeldsen, H. 2003, PASA, 20, 203

\bibitem[Benomaer et al.(2010)]{ben10}
Benomar, O., Baudin, F., Marques, J. P., Goupil, M. J., Lebreton, Y., Deheuvels, S., 2010,
AN 331, 956

\bibitem[Borucki et al.(2010)]{borucki10} Borucki, W. J., Koch, D. G.,
Basri, G., 2010, Sci, 327, 977

\bibitem[Brown et al.(2011)]{br11}
Brown, T. M., Latham, D. W., Everett, M. E., Esquerdo, G. A. 2011, ApJ, in press (arXiv:102.0342)

\bibitem[Cassagrande et al.(2010)]{cas10}
Casagrande, L, Ramirez, I., Melendez, J, Bessel, M., Asplund, M., 2010, A\&A, 512 A54

\bibitem[Chaplin et al.(2008)]{chaplin08}
Chaplin, W. J., Houdek, G., Appourchaux, T., Elsworth, Y., New, R.
\& Toutain, T. 2008, A\&A, 485, 813

\bibitem[Chaplin et al.(2010)]{chaplin10} Chaplin, W. J., Appourchaux,
  T., Elsworth, Y., et al., 2010, ApJ, 713, L169

\bibitem[Chaplin et al.(2011)]{cha11} Chaplin, W. J., Kjeldsen, H.,
Christensen-Dalsgaard, J., Basu, S., Miglio, A. et al. 2011, Science,
332, 213

\bibitem[Christensen-Dalsgaard(1993)]{christensen93}
Christensen-Dalsgaard, J. 1993, in
    ASP. Conf. Ser. 42, Proc. GONG 1992, Seismic Investigation of
    the Sun and Stars, ed. T. M. Brown (San Francisco, CA: ASP), 347

\bibitem[Deheuvels \&Michel(2011)]{sd11}
Deheuvels, S., Michel, E. 2011, A\&A, in press (arXiv:1109.1191) 

\bibitem[Demarque et al.(2004)]{demarque04}
Demarque, P., Woo, J. -H., Kim, Y. -C.,
    \& Yi, S. K. 2004, ApJS, 155, 667

\bibitem[Demarque et al.(2008)]{demarque08}
Demarque, P., Guenther, D. B., Li, L. H., Mazumdar, A. \& Straka, C.
W. 2008, \apss, 316, 311

\bibitem[Dotter et al.(2008)]{dotter08}
Dotter, A., Chaboyer, B., Jevremovic, D. et al. 2008, \apj, 178, 89

\bibitem[Gai et al.(2011)]{gai11}
Gai, N., Basu, S., Chaplin, W. J., Elsworth, Y. 2011, ApJ, 730, 63

\bibitem[Gilliland et al.(2010)]{gill10} Gilliland, R. L., Brown,
  T. M., Christensen-Dalsgaard, J., et al., PASP, 2010, 122, 131

\bibitem[Girardi et al.(2000)]{girardi00} 
Girardi, L., Bressan, A.,
    Bertelli, G \& Chiosi C., 2000, A\&ASS, 141, 371

\bibitem[Grevesse \& Noels(1993)]{gre93}
Grevesse, N., Noels, A. 1993, in
{Origin and Evolution of the Elements}, eds., N. Prantzos,
E. Vangioni-Flam, M. Cass\`e, Cambridge Univ.

\bibitem[Grevesse \& Sauval(1998)]{gs98}
 Grevesse, N.,  Sauval, A.J.,  1998, Sp. Sc. Rev., 85, 161

\bibitem[Grevesse et al.(2010)]{gre10}
Grevesse, N., Asplund, M., Sauval, A. J., Scott, P. 2010, Ap\&SS, 328, 179

\bibitem[Kjeldsen \& Bedding(1995)]{kb95}
Kjeldsen, H., Bedding, T. R. 1995, A\&A, 293, 87

\bibitem[Krishna Swamy(1966)]{krs66}
Krishna Swamy, K. S. 1966, ApJ, 147, 174

\bibitem[Metcalfe et al.(2010)]{met10} Metcalfe, T. S., Monteiro,
M. J. P. F. G., Thompson, M. J., Molenda-\.Zakowicz, J., Appourchaux,
T., et al., 2010, ApJ, 723, 1583

\bibitem[Marigo et al.(2008)]{Marigo08}
Marigo, P., Girardi, L.,
    Bressan, A. et al. 2008, A\&A, 482, 883

\bibitem[Michel et al. (2008)]{michel08} Michel, E., Baglin, A.,
Auvergne, M., et al., 2008, Sci, 322, 558

\bibitem[Pinsonneault et al.(2011)]{Pin11} Pinsonneault, M., An, D.,
Bruntt, H., Molenda-\'Zakowicz, J., Metcalfe, T., Chaplin, W. J.,
2011, ApJ, submitted

\bibitem[Stello et al.(2009)]{ds09} Stello, D., Chaplin, W. J., Basu,
S., Elsworth, Y., Bedding, T. R. 2009, MNRAS, 400, L80

\bibitem[Ulrich(1986)]{rl96}
Ulrich R. K., 1986, ApJ, 306, L37

\bibitem[Takeda et al.(2007)]{ta07}
Takeda, G., Ford, E. B., Sills, A., Rasio, F.A., FIscher, D.A., Valenti, J.A. 2007, ApJS, 168, 297 

\bibitem[Vernazza et al.(1973)]{ver73}
Vernazza, J. E., Avrett, E. H., Loeser, R. 1973, \apj 184, 605

\bibitem[Verner et al.(2011)]{gav11}
Verner, G. A., Chaplin, W. J., Basu, S., Brown, T. M., Hekker, S., et al. 2011, ApJ, 738, L28

\end{thebibliography}
\end{document}